\documentclass[twocolumn,showpacs,amsmath,amssymb]{revtex4}

\usepackage{graphicx}   
\usepackage{dcolumn}   
\usepackage{bm}            

\newcommand{\R}{\mbox{R}}
\newcommand{\Rhat}{\hat{\mbox{R}}}
\newcommand{\Ra}{\mbox{Ra}}
\newcommand{\Nu}{\mbox{Nu}}
\newcommand{\Pran}{\mbox{Pr}}

\def\hat{\widehat}
\def\lap{\Delta}

\def\u{\mbox{\boldmath $u$}}

\begin{document}

\title{A Comparison of Turbulent Thermal Convection Between Conditions of Constant Temperature and Constant Flux}
\author{Hans Johnston}
\affiliation{Department of Mathematics \& Statistics  \\
             University of Massachusetts, Amherst, MA 01003-9305}
\author{Charles R. Doering} 
\affiliation{Departments of Mathematics \& Physics, and Center for the Study of Complex Systems\\
              University of Michigan, Ann Arbor, MI 48109-1043}
\date{\today}

\begin{abstract}
We report the results of  high resolution direct numerical simulations of two-dimensional Rayleigh-B\'enard
convection for Rayleigh numbers up to $\Ra=10^{10}$ in order to study the influence of temperature boundary conditions on turbulent heat transport.
Specifically,  we considered the extreme cases of fixed heat  flux (where the top and bottom boundaries are poor thermal conductors) and fixed temperature (perfectly conducting boundaries).  
Both cases display identical heat transport at high Rayleigh numbers fitting a power law $\Nu \approx 0.138 \times \Ra^{.285}$ with a scaling exponent  indistinguishable from $2/7 = .2857\dots$ above $\Ra = 10^{7}$. 
The overall flow dynamics for both scenarios,  in particular the time averaged 
temperature profiles, are also indistinguishable at the highest Rayleigh numbers.
The findings are compared and contrasted with results of recent three-dimensional simulations.

\end{abstract}


\maketitle

\section{Introduction}

Convection refers to phenomena where spatial inhomogeneities in an advected scalar field drive a fluid flow which in turn transports the scalar.
Convection of various sorts (thermal, compositional, double-diffusive) plays a fundamental role in a wide range of geophysical, astrophysical and engineering applications.
Transport properties of convective flows are of utmost interest and are the focus of scientific efforts worldwide.
Rayleigh-B\'enard convection, the buoyancy driven flow in a fluid layer heated from below, is one of the fundamental paradigms of nonlinear physics, complex dynamics and pattern formation \cite{Kadanoff2001}.
Despite the great deal of effort that has been devoted to it, however, the bulk transport properties of turbulent Rayleigh-B\'enard convection still present challenges for theory, simulation and experiment.

The Rayleigh number $\Ra$ is a ratio of driving due to buoyancy resulting from a vertical temperature gradient, to damping due to the fluid's viscosity and thermal diffusion.  
The enhancement of vertical heat transport by the convectively driven flow is measured by the dimensionless Nusselt number $\mbox{Nu}$. 
The Prandtl number $\mbox{Pr} = \nu/\kappa$, is the ratio of the fluid's kinematic viscosity to its thermal diffusivity.
The goal of many experiments, simulations, theories and analyses is to  discern the behavior of $\mbox{Nu}$ as a function of $\mbox{Ra}$ and $\mbox{Pr}$ and geometric structure (often aspect ratio) of the domain.
When the Prandtl number and aspect ratio are fixed, the Rayleigh number enjoys the status of the control parameter.
Bulk transport in convective turbulence is an open problem: there is still no universally accepted theoretical expectation for what the  asymptotic high Rayleigh number $\mbox{Nu}$-$\mbox{Ra}$ relationship should be \cite{Kraichnan,Spiegel1971,ChicagoJFM,SS90,Yakhot92,GrossmannLohse}, and the state of affairs experimentally is unresolved and even somewhat controversial \cite{Roche,Libchaber87,Castaing,NatureMetal,Oregon,NS2003,SantaBarbara1,SantaBarbara2}. 

Analysis has played and continues to play an important role in this problem.
Mathematically reliable limits on convective transport derived from the fundamental model place constraints
on theories invoking uncontrolled approximations or incorporating additional assumptions.
The classic rigorous result for Rayleigh-B\'enard convection between fixed-temperature no-slip plates is the scaling bound $\mbox{Nu} \le c \mbox{Ra}^{1/2}$
where the prefactor $c$ is uniform in $\mbox{Pr}$ for fixed temperature and no-slip velocity boundary conditions \cite{Howard63,dc96}.  
The $\Nu$\,$\sim$\,$\mbox{Ra}^{1/2}$ scaling has been proposed in several theories \cite{Kraichnan,Spiegel1971,GrossmannLohse}, albeit with distinct Prandtl number dependences, while experimental results have suggested that the asymptotic high-$\mbox{Ra}$ exponent is somewhere between $2/7$ and $1/2$ depending on the Prandtl number and, perhaps, other features.
The $1/2$ scaling bound remains the best known rigorous estimate for the arbitrary-$\Pr$ problem, although $\Nu \lesssim (\Ra \times \ln{\Ra})^{1/3}$ holds for sufficiently high $\Pr$ at fixed $\Ra$ \cite{W} and for infinite $\Pr$ uniformly in $\Ra$ \cite{DOR}.

One proposed explanation for discrepancies among experimental results is the effect of finite conductivity of the upper and lower boundaries \cite{CCC2002,Chilla2004}.
Fixed-temperature (Dirichlet) boundary conditions model plates of infinite thermal conductivity while the limit of poorly conducting boundaries corresponds to fixed heat flux (Neumann) boundary conditions.
Intermediate situations are modeled by interpolating (radiation) boundary conditions.
It is well known that the thermal boundary conditions have a significant effect near the convective transition, decreasing the critical Rayleigh number for the onset of convection and shifting the instability to larger scales \cite{HJP1967}.
Indeed, the long-wave nature of the fixed-flux linear instability has been proposed as an important effect for pattern selection in high-$\mbox{Ra}$ turbulent convection in large aspect ratio systems \cite{HPPPS2008}.

Recent direct numerical simulations in a three-dimensional cylindrical cell ($\mbox{Pr}=.7$) indicated that $\Nu$ is supressed for finite conductivity plates above $\mbox{Ra} = 10^9$, and it was suggested that typical experimental conditions with finite conductivity plates are closer to the fixed flux, rather than the fixed temperature case  \cite{VS2008}.
Rigorous heat transport bounds for the fixed flux case were previously derived \cite{flux}, but they scale $\sim$\,$\mbox{Ra} ^{1/2}$ as in the fixed temperature case, well above the simulation results for either boundary conditions.

In this letter we report the results of a high-resolution computational study of the difference between fixed flux and fixed temperature Rayleigh-B\'enard convection in two spatial dimensions.
We restrict attention to two dimensions in order to guarantee full resolution of the boundary layers at high Rayleigh numbers \cite{JD2007}.
Not unexpectedly, the critical Rayleigh number at convective onset and $\Nu$ immediately above onset differ, with the fixed flux Nusselt number exceeding the fixed temperature Nusselt number as anticipated by linear stability analysis \cite{HJP1967}, but the Nusselt numbers and other qualitative and quantitative features of the flows are observed to coincide to a very high degree of accuracy at high Rayleigh numbers in distinction from the computational results reported for three dimensions.

\section{Mathematical Model and Numerical Scheme}

\begin{figure}
\includegraphics[scale=0.4]{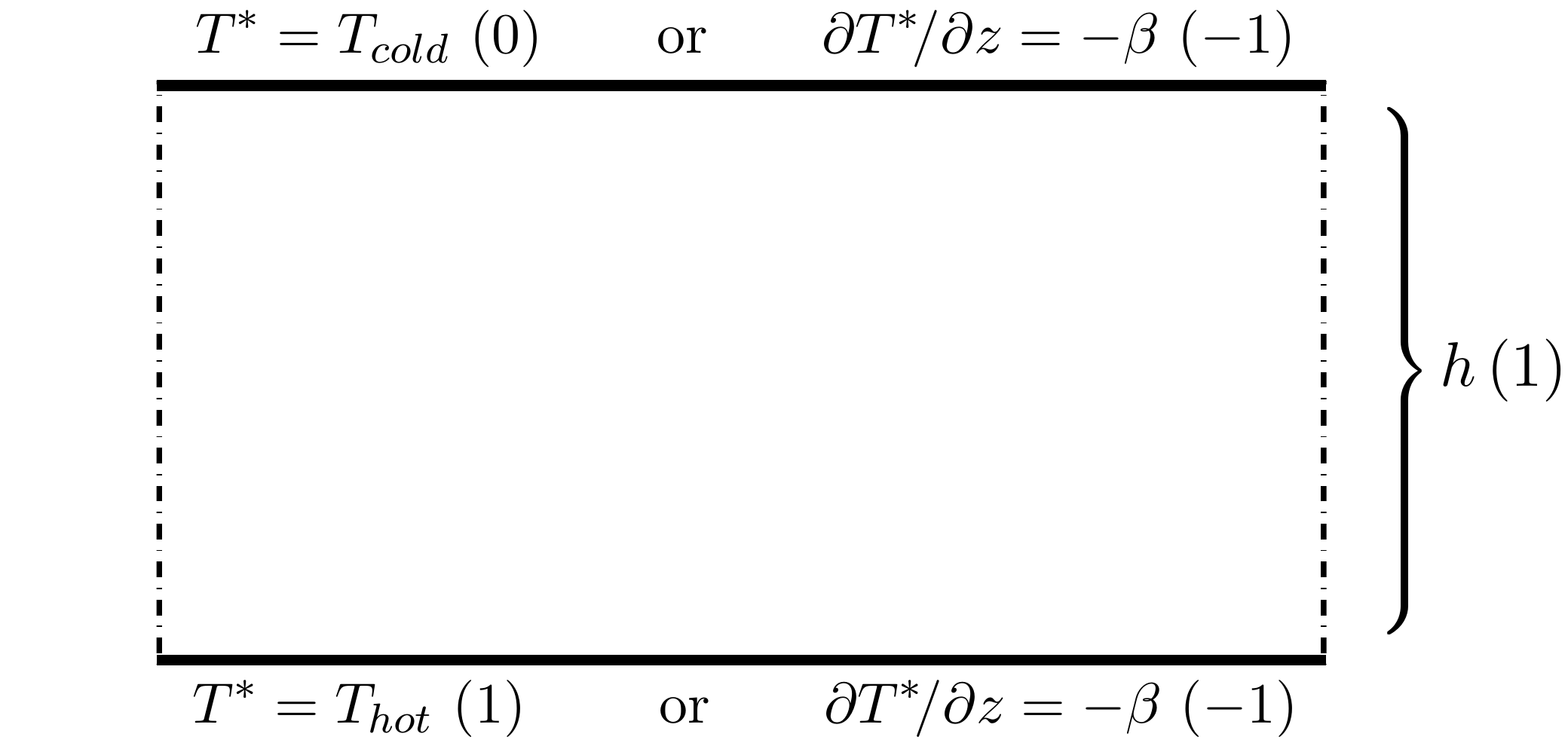}
\caption{2D convection cell, periodic in the horizontal direction, indicating the different temperature boundary conditions considered: fixed temperature and fixed heat flux (non-dimensional values appear in parentheses).}
\label{fig:conv_cell}
\end{figure}

The flow is modeled by the Boussinessq approximation for a unit density incompressible fluid in a horizontally periodic two-dimensional domain of height $h$ with no-slip velocity boundary conditions at the top and bottom plates (Fig.~\ref{fig:conv_cell}).
After a change of variables depending on the temperature boundary conditions, the governing equations are
\begin{eqnarray} 
T_t + (\u \cdot \nabla) T =  {(\Pran \R)}^{-1/2}  \lap T, \label{VS1}
\\[2pt] 
\partial_t \omega + (\u \cdot \nabla) \omega  = {(\Pran/\R)}^{1/2}  \lap \omega - \partial T / \partial x, \label{VS2}
\\
\lap \psi =  \omega , \quad \psi \vert_{z=0,1} = 0, \quad {\partial\psi/\partial z}\vert_{z=0,1} =0, \label{VS3} 
\end{eqnarray} 
where $T$ the temperature, $\u = {\bf i} u + {\bf k} w= {\bf i}  \partial_z  \psi - {\bf k}  \partial_x \psi$ is the velocity field, and $\omega = \partial_z u - \partial_x w$ is the vorticty.
The control parameter $R$  in the measure of the imposed thermal forcing whose definition depends on the boundary conditions:

For the case of fixed temperature at each boundary the space, time, and temperature scales are, respectively, $h$,  ${(h/\alpha g \delta T^*)}^{1/2}$,  and $\delta T^* =  T_{hot}-T_{cold}$, the dimensional temperature drop, where $g$ is the acceleration due to gravity and $\alpha$ is the thermal expansion coefficient of the fluid.
Then the thermal forcing parameter $R$ in (\ref{VS1}) and (\ref{VS2}) is the usual Rayleigh number 
\begin{equation}
\R = \Ra =    \alpha g  \delta T^* h^3/\nu\kappa \label{FT}
\end{equation}
with temperature boundary conditions $T \vert_{z=0} = 1$ and $T \vert_{z=1} = 0$.
For the case of an imposed vertical  fixed heat flux $\sim$\,$\beta$, the space, time, and temperature scales are, respectively, $h$,  ${(\alpha g \beta)}^{-1/2}$, and $\beta h$.  Then in place of (\ref{FT}),
\begin{equation}
\R = \Rhat = \alpha g \beta h^4/\nu\kappa\label{FF}
\end{equation}
with $\partial T/ \partial z \vert_{z=0,1} = -1$.

Denote the space-time average of a function $f({\bf x},t)$ by $\langle \, f \, \rangle$.  
Then for the fixed-temperature boundary conditions the Nusselt number is
\begin{equation}
\Nu = 1+{(\Pran\Ra)}^{1/2} {\langle w T\rangle },
\end{equation}
while for the fixed-flux case it is
\begin{equation}
\Nu =  \left(1-{(\Pran\Rhat)}^{1/2} {\langle w T\rangle} \right)^{-1}.
\end{equation}
Note that  $\Rhat = \Ra \Nu$ \cite{flux,VS2008}.

The numerical scheme used for simulating (\ref{VS1})-(\ref{VS3})  is a 
Fourier-Chebyshev spectral collocation method in space with classical fourth order Runge-Kutta  
for the time stepping \cite{WJL}.   Computation of the momentum equation (\ref{VS2})  and the
kinematic constraint  (\ref{VS3}) are decoupled through use of a high order local formula for the 
vorticity at the boundary, derived from the Neumann boundary condition 
${\partial\psi/\partial z}\vert_{z=0,1} =0$ for the stream function.   The  Dirichlet boundary condition
$\psi \vert_{z=0,1} = 0$ is imposed in the solution of the elliptic system (\ref{VS3}), 
which is solved by the matrix-diagonalization procedure \cite{GottLuts,EhrenPeyret}.  
To ensure benchmark quality simulations, the grid sizes were chosen with a minimum 
of eight grid points in the thermal boundary layer, defined as the distance from the boundary 
at which the extrapolation of the linear portion of the mean profile at the boundary equals
the central mean temperature \cite{BTL1994,VS2008}.   Achieving this resolution  over the
full range of $\Ra$, while maintaining computational efficiency and accuracy,
was made possible through the use of the Kosloff \& Tal-Ezer mapping \cite{KosTal} applied to the
Chebyshev points and a high order exponential filter to control aliasing errors. 

\section{Results and Discussion}

\begin{figure}
\includegraphics[scale=0.55]{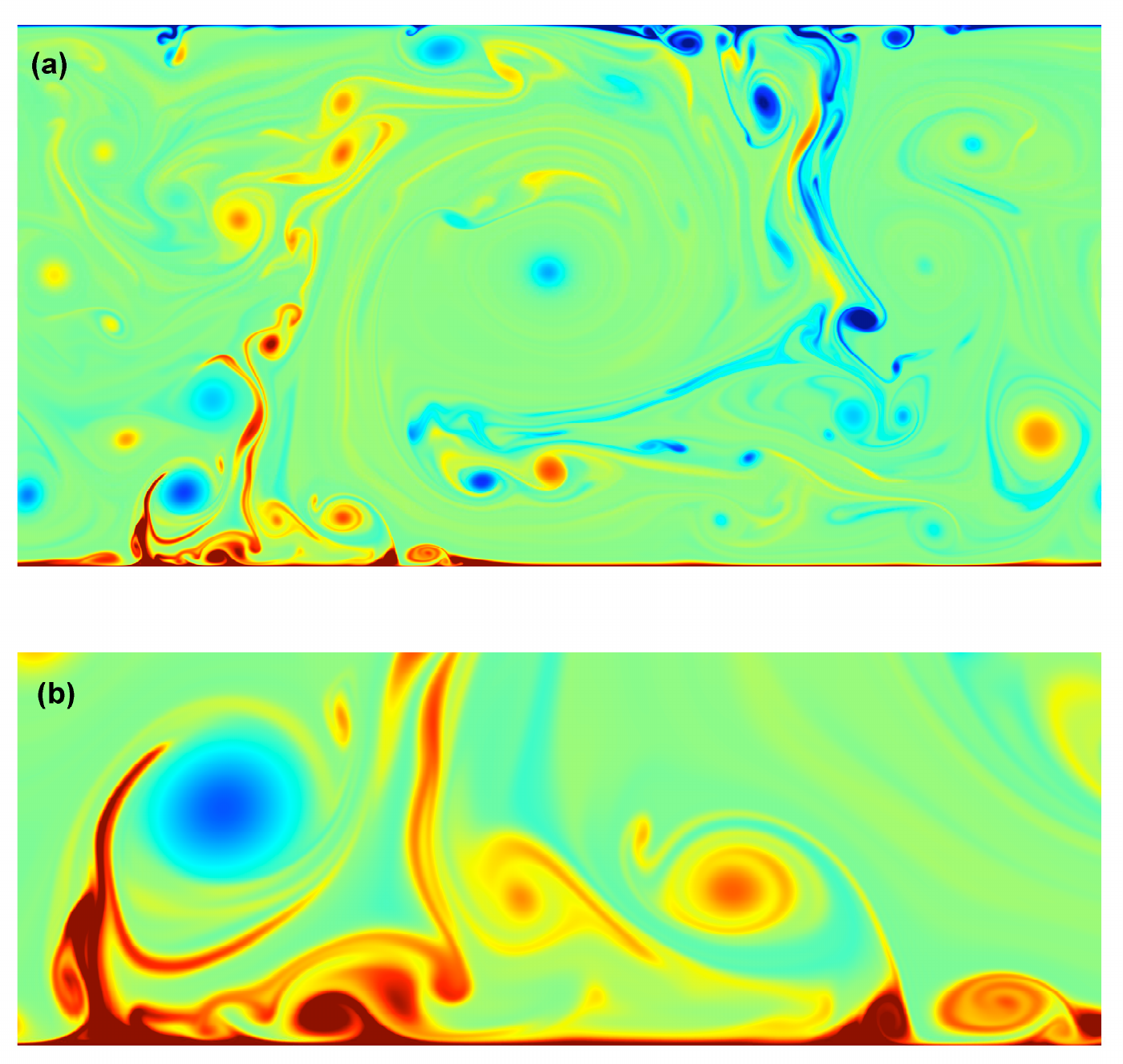}
\caption{Temperature snapshot for fixed flux convection at $\Ra = 1.05\times10^{10}$
($\Rhat = 1.07\times 10^{12}$): (a) full domain, (b) zoom image of boundary layer to illustrate resolution.}
\label{fig:Ra10bil_ff}
\end{figure}

\begin{figure}
\includegraphics[scale=0.45]{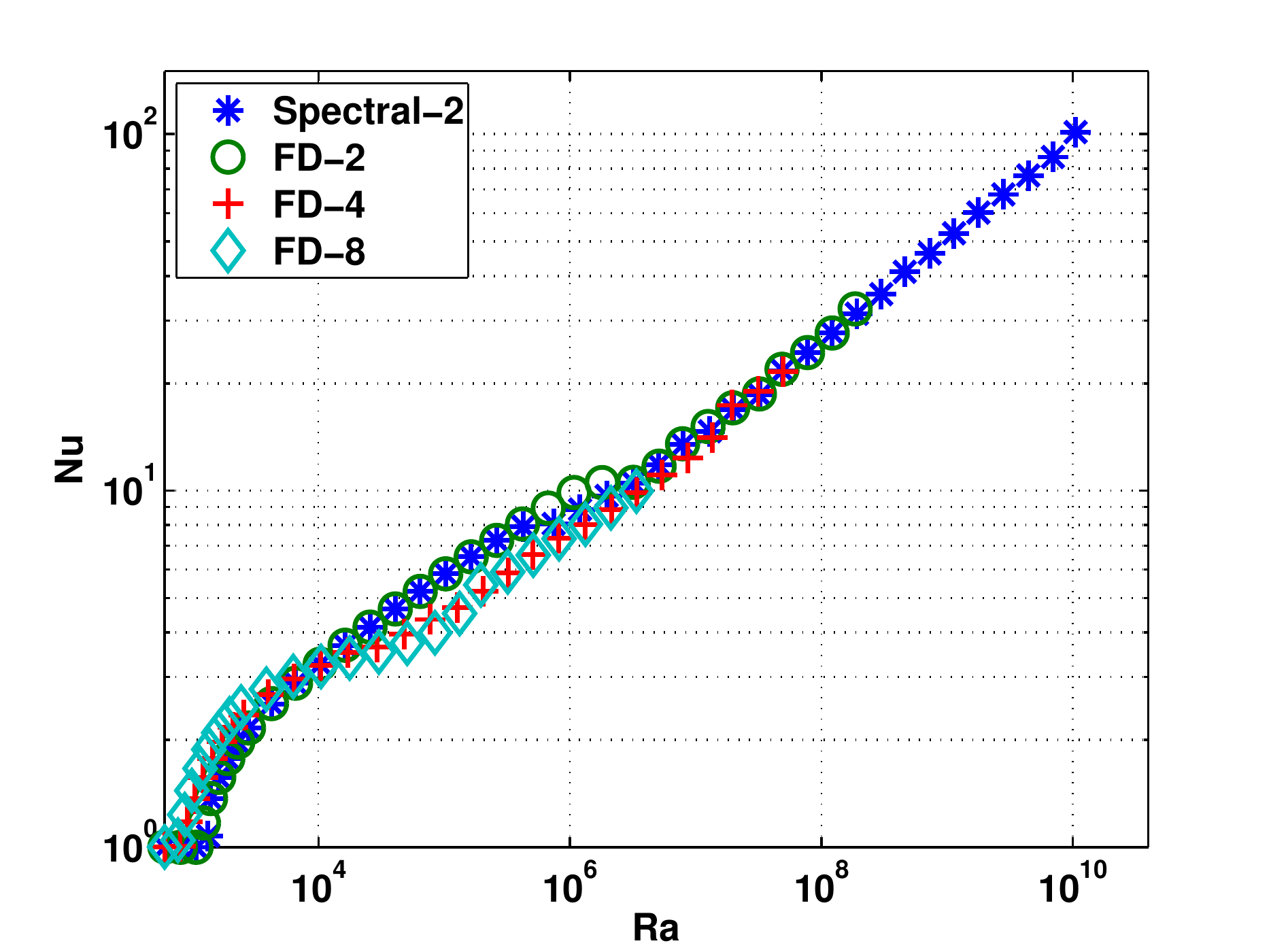}
\caption{$\Ra$-$\Nu$ data for fixed-flux simulations.
The spectral code was used for an aspect ratio 2 cell (Spectral-2),
and the fourth order finite difference method 
\cite{LWJ} was used for cells of aspect ratio 2 (FD-2), 4 (FD-4), and 8 (FD-8).}
\label{fig:ranu_ff}
\end{figure}

All results reported here are for $\Pr = 1$.
Simulations were first performed for the fixed heat flux case in a cell of aspect ratio $2$.
Beginning at $\Rhat=600$---below onset of convection---$\Rhat$ was increased step-by-step by a constant factor, allowing at the flow to settle into a steady or statistically stationary dynamical state before proceeding.  
A snapshot of the temperature field at the highest Rayleigh number is shown in  Fig.~\ref{fig:Ra10bil_ff}.
For each value of $\Rhat$ the  Nusselt number $\Nu$ was measured and the effective Rayleigh number recovered from  $\Ra = \Rhat/\Nu$.  

In order to validate the fixed flux simulations before moving to the fixed temperature
case, the numerical
experiment was repeated using a fourth order finite difference method \cite{LWJ} following the same
protocol for cells of aspect ratio 2, 4, and 8, over a more restricted range of $\Ra$ in order to
achieve the same numerical accuracy as in the spectral simulations.  
Fig.~\ref{fig:ranu_ff} shows all these fixed-flux $\Ra$-$\Nu$ data sets,
noting that the simulation times were sufficiently long to ensure that uncertainties in the $\Nu$ 
measurements are within the size of plot symbols. 
For fixed flux Rayleigh-B\'enard convection the critical value of $\Rhat$ at onset depends on (decreases with) the aspect ratio, and the wavelength at onset 
is set by the horizontal---rather than the vertical---scale of the domain \cite{HJP1967}.
Hence the initial pair of convection rolls is as wide as the aspect ratio permitted.
But at each aspect ratio, as the Rayleigh number was increased the roles eventually became unstable, turbulent convection sets in, and the flow was observed to organize itself into pairs of turbulent aspect-ratio 2 cells like those in Fig.~\ref{fig:Ra10bil_ff}.
These observations, together with the data in Fig.~\ref{fig:ranu_ff}, suggest that two-dimensional fixed-flux convective turbulence is independent of the aspect ratio at asymptotically high Rayleigh numbers.

\begin{figure}
\includegraphics[scale=0.45]{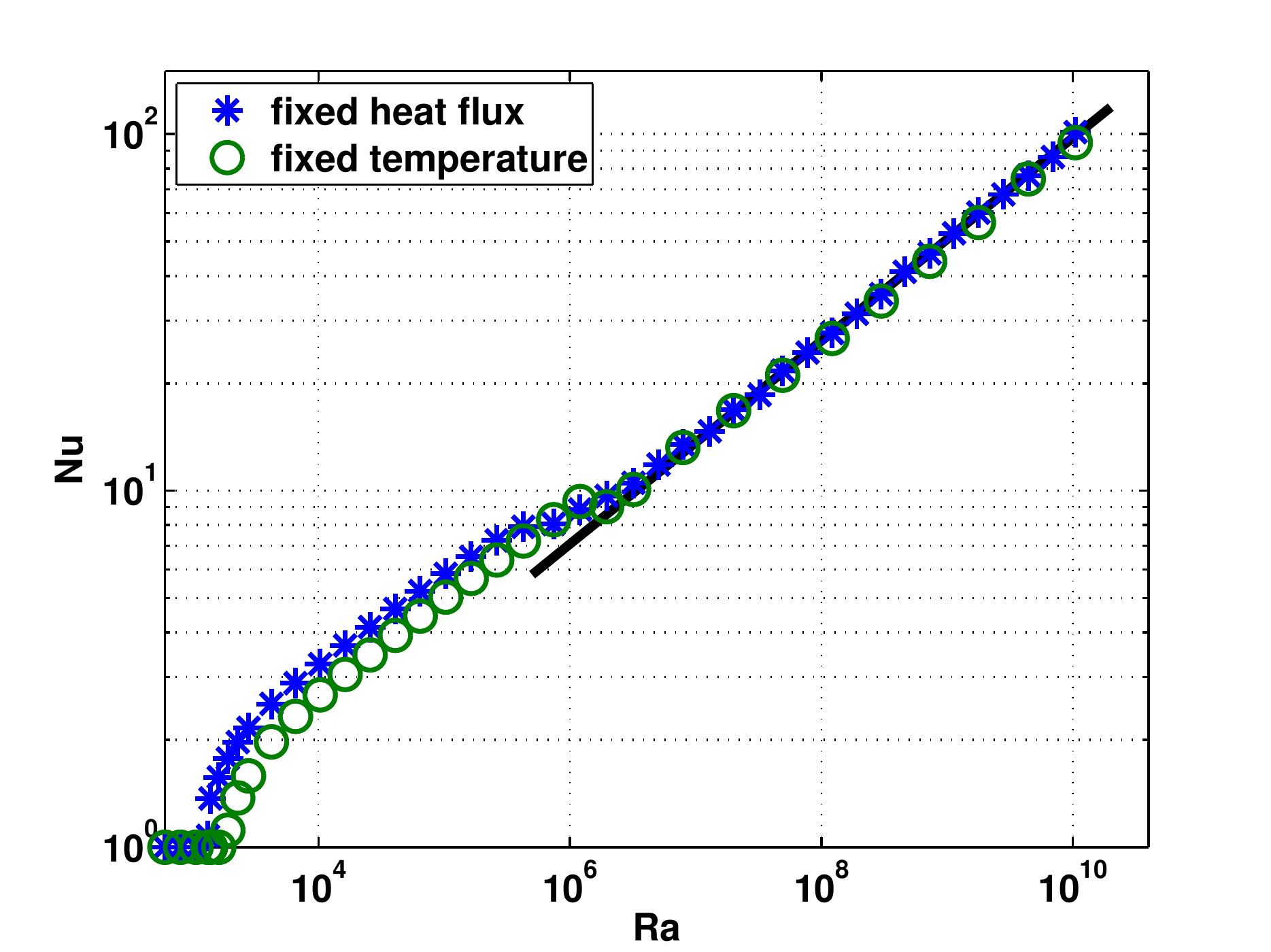}
\caption{Comparison of heat transport data for fixed-flux and fixed-temperature simulations in a cell of aspect ratio 2.}
\label{fig:ranu_ffft}
\end{figure}

Following the same protocol, simulations with fixed temperatures at the boundaries were carried 
out for the same values of $\Ra$ in an aspect ratio 2 cell.  
Fig.~\ref{fig:ranu_ffft} shows the $\Ra$-$\Nu$ data for both scenarios.  It is observed that the data are the same, within the simulations' uncertainties, at high Rayleigh numbers.
A fit of the eight highest $\Ra$ data points generated with the fixed flux boundary conditions yields 
$\Nu \approx 0.138 \times \Ra^{.285}$ with a scaling exponent  indistinguishable from $2/7 = .2857\dots$.
This $\Nu$-$\Ra$ relationship is in remarkably close agreement with the latest high-precision three-dimensional simulations for fixed-temperature conditions in a cylinder with insulating walls \cite{ES2008} which produced  $\Nu = 0.175 \times \Ra^{.283}$ for $10^7 \le \Ra \le 10^9$ at $\Pr = .7$.

\begin{figure}
\includegraphics[scale=0.44]{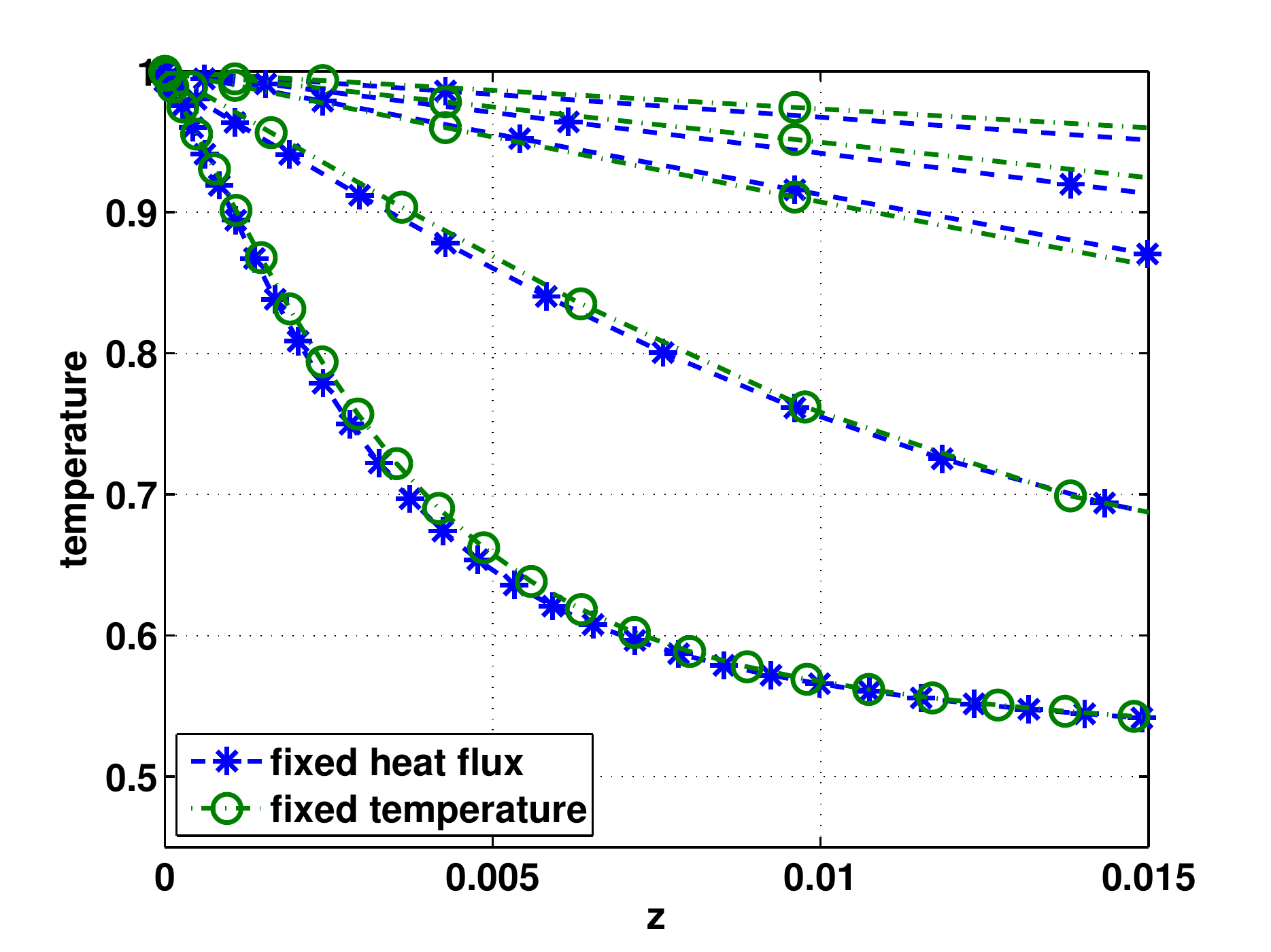}
\caption{Time averaged temperature profiles (from right to left): $Ra\!=\!1.03\times10^4$, $1.03\times10^5$, $1.21\times10^6$, $1.22\times10^8$, and $1.05\times10^{10}$.  The discrete data indicate the spatial discretization used to resolve the boundary layers.}
\label{typrof}
\end{figure}

The quantitative correspondence between fixed-flux and fixed-temperature (two-dimensional) turbulent convection is not limited to the bulk heat flux.  
The mean temperature profiles are observed to converge at high Rayleigh numbers as well.
Fig.~\ref{typrof} shows horizontally and temporally averaged temperature profiles for the two scenarios.
At the highest Rayleigh numbers they are effectively indistinguishable.
We have not systematically compared other statistical quantities, but we have noted striking similarities between the large scale dynamics at fixed flux and fixed  temperature \cite{movies}.

The $\Nu \sim \Ra^{2/7}$ relationship was previously observed with fixed temperature boundaries in two-dimensional simulations  \cite{werne}, and in three-dimensional simulations both with rotation \cite{julien} and without \cite{kerr}, albeit at significantly lower Rayleigh numbers.
Recent direct numerical simulations at much higher Rayleigh numbers (and $\Pr=.7$) reported $\Nu \sim \Ra^{1/3}$ over nearly four decades up to $\Ra = 10^{14}$ in a three-dimensional cylindrical cell of aspect ratio $1/2$ \cite{amati}.
It remains to be seen whether the heat transport in two-dimensional Rayleigh-B\'enard convection ever deviates from the $\Nu$\,$\sim$\,$\Ra^{2/7}$ scaling at higher Rayleigh numbers.

In summary, the high-resolution simulation results reported here suggest that the plate conductivity plays no significant role in the $\Nu$\,$\sim$\,$\Ra^{2/7}$ transport law at asymptotically high $\Ra$ in two dimensions with periodic side conditions at $\Pr=1$.
However, it remains to be determined how various combinations of geometry, side-wall conditions, plate conductivity and the spatial dimension affect the bulk transport in high-resolution simulations at asymptotically high Rayleigh numbers.

{\bf Acknowledgements:} This work was supported by NSF PHY-0555324.  We thank K. Julien, J. Schumacher, E. A. Spiegel, and J. Werne for helpful discussions.

\eject


\begin{thebibliography}{99} 

\bibitem{Kadanoff2001}  L.P. Kadanoff, Physics Today  {\bf 54}, 34 (2001).

\bibitem{Kraichnan} R. H. Kraichnan, Phys. Fluids {\bf 5}, 1374 (1962).

\bibitem{Spiegel1971} E.A. Spiegel, Ann. Rev. Ast. \& Ast.  {\bf 9}, 323 (1971).

\bibitem{ChicagoJFM} B. Castaing {\it et al}, J. Fluid Mech. {\bf 204}, 1 (1989).

\bibitem{SS90} B. Shraiman \& E. Siggia, Phys. Rev. A {\bf 42}, 3650 (1990).

\bibitem{Yakhot92} V. Yakhot, Phys. Rev. Lett. {\bf 69}, 769 (1992).

\bibitem{GrossmannLohse} S. Grossmann \& D. Lohse, J. Fluid Mech.  {\bf 407}, 27 (2000).

\bibitem{Roche} P. Roche {\it et al}, Phys. Rev. E  {\bf 63}, 045303 (2001).

\bibitem{Libchaber87} F. Heslot {\it et al}, Phys. Rev. A {\bf 36}, 5870 (1987).

\bibitem{Castaing} X. Chavanne {\it et al}, Phys. Rev. Lett. {\bf 79}, 3648 (1997).

\bibitem{NatureMetal} J. Glazier {\it et al}, Nature  {\bf 398}, 294 (1999).

\bibitem{Oregon} J. Niemela {\it et al}, Nature {\bf 404} 837 (2000).

\bibitem{NS2003} J. Niemela \& K.R. Sreenivasan, J. Fluid Mech.  {\bf 481}, 355 (2003).

\bibitem{SantaBarbara1} A. Nikolaenko {\it et al},  J. Fluid Mech. {\bf 523}, 251 (2005).

\bibitem{SantaBarbara2} D. Funfschilling {\it et al},  J. Fluid Mech.  {\bf 536}, 145 (2005).

\bibitem{Howard63} L.N. Howard, J. Fluid. Mech.  {\bf 17}, 405 (1963).

\bibitem{dc96} C.R. Doering \& P. Constantin, Phys. Rev. E {\bf 53}, 5957 (1996).

\bibitem{W} X.M. Wang, Commun. Pure \& Appl. Math. {\bf 61} 789 (2008).

\bibitem{DOR} C. R. Doering, F. Otto \& M. Reznikoff, J. Fluid Mech. {\bf 560}, 229 (2006).

\bibitem{CCC2002} S. Chaumat  {\it et al}, in {\it Advances in Turbulence IX}, ed. I. P. Castro \& P. Hancock (2002).

\bibitem{Chilla2004} F. Chilla {\it et al}, Phys. Fluids  {\bf 16}, 2452 (2004).

\bibitem{HJP1967} D. T. J. Hurle {\it et al}, Proc. Roy. Soc. Lon., Ser. A {\bf 296}, 469 (1967).


\bibitem{HPPPS2008} J. von Hardenberg {\it et al}, Phys. Lett. A {\bf 372}, 2223 (2008).

\bibitem{VS2008} R. Verzicco \& K. R. Sreenivasan, J. Fluid Mech. {\bf 595}, 203 (2008).

\bibitem{flux} J. Otero {\it et al},  J. Fluid Mech.  {\bf 473}, 191 (2002).

\bibitem{JD2007} H. Johnston and C. R. Doering, Chaos {\bf 17}, 041103 (2007).

\bibitem{WJL} C.\ Wang, H.\ Johnston \&  J.-G.\ Liu, unpublished. 

\bibitem{EhrenPeyret}  U. Ehrenstein \& R. Peyret, Intern. J. Numer. Methods Fluids, {\bf 9} 427  (1987).   

\bibitem{GottLuts}  D.\ Gottlieb \& L.\ Lustman, SIAM J.\ Numer. Anal. {\bf 20}, 909 (1983).                                               

\bibitem{BTL1994} A. Belmonte  {\it et al}, Phys. Rev. E {\bf 50}, 269 (1994).

\bibitem{KosTal}  D.\ Kosloff \& H.\ Tal-Ezer,  J. Comput. Phys. {\bf 104}, 457 (1993). 

\bibitem{LWJ}  J.-G.\ Liu, C.\ Wang \& H.\ Johnston, J. Sci. Comp. {\bf 18}, 253 (2003).

\bibitem{ES2008} M. S. Emran \& J. Schumacher, J. Fluid Mech. {\bf 611}, 13 (2008).

\bibitem{movies} See additional images and movies of the simulations at 
\url{www.math.umass.edu/~johnston/RayBen.html}

\bibitem{werne} E. E. DeLuca {\it et al}, Phys. Rev. Lett. {\bf 64}, 2370 (1990).

\bibitem{julien} K. Julien  {\it et al}, Phys. Rev. E {\bf 53}, R5557 (1996).

\bibitem{kerr} R. M. Kerr, J. Fluid Mech. {\bf 59}, 139 (1996).

\bibitem{amati} G. Amati {\it et al}, Phys. Fluids {\bf 17}, 121701 (2005).


\end{thebibliography}
\end{document}